\documentclass{article}
\usepackage{a4,amsmath,amsfonts}
\usepackage{graphics}
\usepackage{graphicx}
\usepackage{fancyhdr}
\usepackage{xcolor} 
\usepackage[all,cmtip]{xy}
\usepackage{amsmath}
\usepackage{amsthm}
\usepackage{amssymb}

\usepackage{setspace}

\allowdisplaybreaks

\begin{document}
\begin{center}
\textbf{\huge Two Dimensional Plane, Modified Symplectic Structure and Quantization}
\vskip 0.3cm \Large Mohd Faudzi Umar$^{a,b}$,  Nurisya Mohd Shah$^{b,c}$ and Hishamuddin Zainuddin$^{b,c}$
\\[0.5cm]
\begin{spacing}{1}
\small
\vskip 0.1cm $^a$ \textit{Faculty of Science and Mathematics, Universiti Pendidikan Sultan Idris,}\\
\textit{35900 Tanjung Malim, Perak.}\\
\vskip 0.1cm $^b$ \textit{Institute for Mathematical Research (INSPEM), Universiti Putra Malaysia,}\\
 \textit{43400 Serdang, Selangor.}\\
 \vskip 0.1cm $^c$ \textit{Department of Physics, Faculty of Science, Universiti Putra Malaysia, }\\
 \textit{43400 Serdang, Selangor.}\\
\vskip 0.2cm
\end{spacing}
\end{center}

\noindent \textbf{Abstract.} \ Noncommutative quantum mechanics on the plane has been widely studied in the literature. Here, we consider the problem using Isham\textsc{\char13}s canonical group quantization scheme for which the primary object is the symmetry group that underlies the phase space. The noncommutativity of the configuration space coordinates requires us to introduce the ‘noncommutative term’ in the symplectic structure of the system. This modified symplectic structure will modify the group acting on the configuration space from abelian $\mathbb{R}^2$ to a nonabelian one. As a result, the canonical group obtained is a deformed Heisenberg group and the “canonical” commutation relation (CCR) corresponds to what is usually found in noncommutative quantum mechanics.
\\
\noindent {\bf Keywords:} \ Noncommutative quantum mechanics, canonical group quantization, symplectic structure.
\begin{center}
\textbf{INTRODUCTION}
\end{center}
Group-theoretic or canonical group quantization, proposed by C. J. Isham$^1$, is the quantization technique that is geometrical in nature with the group structure as a main ingredient in the scheme. The step starts by identifying the canonical group describing the symmetries of the phase space of the system. Then the generators of the group corresponding to the classical observables will be quantized by finding their irreducible unitary representations. The conventional quantum mechanics can be obtained also when we apply the scheme on the linear phase space, $\mathbb{R}^2$, where the following well-known canonical commutation relation (CCR) will be obtained,
\begin{align}
[\hat{q}^i, \hat{q}^j]=0, \qquad [\hat{p}_i, \hat{p}_j]=0,\qquad [\hat{q}^i, \hat{p}_i]=i\hbar,
\end{align}
where $i,j=1,2$ for a two-dimensional plane as the configuration space and $\hbar$ is the reduced Planck constant. Self-adjoint operators corresponding to the coordinates to the appropriate canonical group which is the Heisenberg group.

In the early works of this quantization scheme, it was proposed for the quantization of nonlinear systems such as gravity$^2$, string on tori$^3$ and particle on torus in a constant magnetic field$^4$ background. Symmetry of the phase space has played a main role in this quantization besides the geometrical structures which are often used in the initial prequantization stage – namely in first identifying canonical group. The procedure can then be extended to the noncommuting position coordinates, and the generalized CCR will be obtained as in \textit{Noncommutative quantum mechanics} (NCQM) in the literature (for more details, see Ref. 5). The generalized CCR requires noncommuting position operators which motivates us to modify the framework$^{1,2,3,4}$ to accommodate this problem. Moreover this situation also gives another natural question: what is the symmetry group of this new phase space with noncommuting coordinates? 

The purpose of this work is to study the NCQM via group theoretic approach. This is done by introducing an additional noncommutative term in the symplectic structure of the system. The discussion starts with the framework, namely canonical group quantization and the following section is the modified symplectic algebra. The section ends with finding the unitary irreducible representations by obtaining the CCR of the NCQM.
\\
\begin{center}
	\textbf{CANONICAL GROUP QUANTIZATION}
\end{center}
Canonical group quantization (CGQ) is a quantization approach that shares the same geometric tools as geometric quantization but takes the cotangent bundle structure of the phase space into account and the canonical group takes the form of a semidirect product. This allows the use of Mackey's theory of induced representations$^9$ for the canonical group. Note that the key idea is the canonical group must describe the symmetries of the phase space i.e. it leaves the symplectic form invariant.  To establish the canonical group for the non-commutative plane, we have to ensure construction of the maps in the following  commutative diagram be satisfied,
\begin{align}\label{2}
\xymatrix{
	0 \ar[r]&\mathbb{R}\ar[r]&C^{\infty}(S,\mathbb{R})\ar[r]^j &HamVF(S)\ar[r]&\mathbb{R}\ar[r]&0
	\\&&&\mathcal{L}(\mathcal{C})\ar@{.>}[lu]^P\ar[u]^{\gamma}}
\end{align}

\noindent The quantization stage is essentially the construction of irreducible unitary representation of the canonical group with its generators being the quantized version of the observables. 

Canonical group quantization has the following steps: Identify a Lie group $\mathcal{G}$, where each element of the Lie algebra $A\in\mathcal{L}(\mathcal{G})$ will induce a vector field via the group action n the phase space. Note that the one-parameter subgroup it generates, $t\mapsto \exp(-tA)\in\mathcal{G}$ induces the vector field
\begin{align}\label{3}
\xi^A=\frac{\partial}{\partial t} se^{-tA}|_{t=0},
\end{align}
where $s$ is a point on the phase space, $\mathcal{S}$. From each Hamiltonian vector field, $\xi^A$, there will be an observable, $f^A\in C^{\infty}(\mathcal{S}, \mathbb{R})$ from the relation
\begin{align}\label{4}
\xi^A\lrcorner\ \omega =df^A\ ,
\end{align}
where $\omega$ is the symplectic form. Alternatively, for each Lie algebra element, there will be an observable under moment map,
\begin{align}\label{5}
P:\mathcal{L}(\mathcal{G})\rightarrow P^A\in C^{\infty}(\mathcal{S}, \mathbb{R}).
\end{align}
 If those maps are made to correspond to each other, we can claim that the Lie group $\mathcal{G}$ is a valid canonical group $\mathcal{C}$. In the final step, one seeks for inequivalent unitary representations of the canonical group  $\mathcal{U}(\mathcal{C})$ to realize inequivalent quantizations of the system. 
\\

\begin{center}
	\textbf{MODIFIED SYMPLECTIC ALGEBRA}
\end{center}
The phase space is given by the cotangent bundle, $T^\ast \mathbb{R}^2\approx\mathbb{R}^2\times \mathbb{R}^2$. The symplectic structure will be modified in order to accommodate the noncommutative positions.  Thus the modified symplectic structure$^8$ is given by
\begin{align}\label{6}
\Omega=dq^i\wedge dp_i+\frac{1}{2}\theta^{ij}dp_i\wedge dp_j,
\end{align}
where $i,j=1,2$, and noncommutative term being, $\omega_{\theta}=\frac{1}{2}\theta^{ij} dp_i˄ dp_j$, with an antisymmetric parameter is anti-symmetric,    $\theta^{ij}=-\theta^{ji}$. In Ref. 8, this is called as dual magnetic field. The natural symplectic structure, $\omega$ will be obtained if parameter $\theta\rightarrow 0$. 

We first attempt the group to be the translation group with elements $(a^i,b_i )\in(\mathbb{R}^2\times \mathbb{R}^2,+)$, and the group action $l_{(a^i,b_i )}$ symplectically and effectively translates points on the phase space $(q^i,p_i )$ by $(q^i+a^i,p_i-b_i)$. Using (\ref{3}) and parametrising the translations by corresponding elements of the algebra $(A^i,B_i)$, the Hamiltonian vector field is given by
\begin{align}\label{7}
\xi^{(A^i, B_i)}=A^i\frac{\partial}{\partial q^i }-B_i\frac{\partial}{\partial p_i}\ .
\end{align}
The contraction between vector field (\ref{7}) and symplectic structure (\ref{6}) gives an exact form (differential of an observable) and the canonical observable reads as
\begin{align}\label{8}
\xi^{(A_i, B_i)}\lrcorner\Omega&= df^{(A_i, B_i)}\ ;\\\label{9}f^{(A_i, B_i)} &= A^ip_i+B_i(q^i-\frac{1}{2}\theta^{ij}p_j)\ .
\end{align}
 Let $\xi^{({A^{\prime}}^i,B^{\prime}_i)}$ be another vector field and performing consecutive contraction procedure will derive a redefined Poisson bracket$^6$:
\begin{align}\label{10}
\left\{f^{(A^i, B_i)}, f^{({A^{\prime}}^i, B^{\prime}_i)}  \right\}= \left(
\frac{\partial f^{(A^i, B_i)}}{\partial q^i}\frac{\partial f^{({A^{\prime}}^i, B^{\prime}_i)}}{\partial p_i}-\frac{\partial f^{(A^i, B_i)}}{\partial p_i}\frac{\partial f^{({A^{\prime}}^i, B^{\prime}_i)}}{\partial q^i}\right)+\frac{1}{2}\theta^{ij}\frac{\partial  f^{(A^i, B_i)}}{\partial q^i }\frac{\partial f^{({A^{\prime}}^i, B^{\prime}_i)}}{\partial q^j}.
\end{align}
The canonical observables (\ref{9}) correspond to the vector fields (\ref{7}) will be redefined then with the extended term $\frac{1}{2}\theta^{ij}\frac{\partial}{\partial q^j}$. Such redefined vector field corresponding to Lie algebra element $(A^i,B_i )$ shows the homomorphism of the $\gamma$ map. From (\ref{9}) the canonical observables are redefined accordingly as the (noncommutative) position,  
${q^{\prime}}^i$ is then extended by the commuting position, $q^i$, while momenta, $p_i$ is unchanged. 
These transformations are also called Bopp shift in literature$^5$:	
\begin{align}\label{11}
q^{i\prime}&=q^i-\frac{1}{2}\theta^{ij}p_j,\\
\label{12} 
p^{\prime}_i&=p_i.
\end{align}
The results in (\ref{11}-\ref{12}) and the contracted relation (\ref{8}) implies the homomorphism of the $j$ map. In particular, the new Poisson brackets between the canonical observables are:
\begin{align}\label{13}
\left\{q^{i\prime}, q^{j\prime}  \right\}= \theta^{ij}, \qquad \left\{p^{\prime}_i, p^{\prime}_j   \right\}=0 , \qquad \left\{q^{i\prime},  p^{\prime}_i   \right\}=1 .
\end{align}
The modified Poisson brackets are anti-symmetric, while Leibiniz rule and Jacobi identity are also satisfied. Furthermore the brackets generalize the Newton’s second law by a noncommutative term (see Ref. 6).

\begin{center}
	\textbf{Canonical Group and Extensions}
\end{center}
Moment map $P$ gives  $P^{(A^i,B_i)}=A^i p^{\prime}_i+B_i {q^{\prime}}^i$ and the homomorphism relation should be given by
\begin{align}\label{14}
\left\{P^{(A^i, B_i)}  , P^{(A^{i\prime}, B^{\prime}_i)} \right\}=P^{[(A^i, B_i)(A^{i\prime}, B^{\prime}_i)]}.
\end{align}
However the relation is not homomorphic since the new Poisson brackets produce two obstructions, 
\begin{align}\label{15}
z_1((A^i, B_i),(A^{i\prime}, B^{\prime}_i))=B_iA^{i\prime}-B^{\prime}_iA^i,
\\
z_1(( B_i,B_j),(B^{\prime}_i, B^{\prime}_j))=\theta^{ij}(B_iB^{\prime}_j-B^{\prime}_iB_j),\label{16}
\end{align}
while Lie bracket on the right hand side of (\ref{14}) is abelian. These obstructions can be removed using two central extension forms such as $z_1=C$ and $z_2=D$. Hence these extensions will now be included in the algebra of $(A^i,B_i,C,D)$, and therefore a redefined Lie bracket in right hand side of (\ref{14}) will be
\begin{align}\label{17}
[(A^i, B_i, C, D), (A^{i\prime}, B^{\prime}_i, C^{\prime}, D^{\prime})]=(0,0,B_iA^{i\prime}-B^{\prime}_iA^i, \theta^{ij}(B_iB^{\prime}_j-B^{\prime}_iB_j) ).
\end{align}
The group will now be extended as well with elements $(a^i,b_i )\rightarrow (a^i,b_i,c,d)$ and the group law being
\begin{align}\label{18}
(a^i, b_i, c, d) (a^{i\prime}, b^{\prime}_i, c^{\prime}, d^{\prime})=(a^i+a^{i\prime}, b_i+b^{\prime}_i, c+c^{\prime}+\frac{1}{2}(b_ia^{i\prime}-b^{\prime}_ia^i), d+d^{\prime}+\frac{1}{2}(b_ib^{\prime}_j-b^{\prime}_ib_j)).
\end{align}
\\

\begin{center}
	\textbf{NONCOMMUTATIVE QUANTUM MECHANICS}
\end{center}
In this section, we briefly attempt to develop the unitary irreducible representations of the canonical group, $\mathcal{C}$. However we have to verify the canonical group of the phase space in the following proposition:
\\

\noindent\textbf{Proposition}
Deformed Heisenberg group, $H^2_{\theta}$ is the canonical group of the phase space with the extended noncommutative term. 
\\

\noindent\textit{Proof.}
To verify the canonical group of the phase space, then the algebras of vector fields and observables can be made to correspond to the Lie algebra using $\gamma$  and $P$ map constructed in (\ref{8}) and (\ref{14}-\ref{17}) respectively. The algebras of the vector fields and observables are related to one another by $j$ map in (\ref{9}-\ref{12}). The group $H^2_{\theta}$ obtained is the centrally extended Heisenberg group $H^2 \rtimes \mathbb{R}$ by the element, $b_i$ and its group law is given in (\ref{18}).
\hfill \( \square \)
\\

Unitary representations of elements of the canonical group, $(a^i,b_i,c,d)\in H^2_{\theta}$ are $U(a^i )=e^{-iA^i\hat{p}^{\prime}_i}$, $V(b_i )=e^{-iB_i \hat{q}^{i\prime}  }$ and $W(c,d)=e^{-i(C\hbar+D\theta)}$, and then the operators satisfy the  following relations,  
\begin{align}\nonumber
U(a^i)U(a^{i\prime})&=U(a^{i\prime})U(a^i),\\
V(b_i)V(b^{\prime}_j)&= V(b^{\prime}_j)V(b_i)e^{i\theta^{ij}B^{\prime}_j B_i}\nonumber,\\
V(b_i)U(a^i)&=U(a^i)V(b_i)e^{i\hbar B_iA^i} \nonumber, \\
U(a^i)W(c,d)   &= W(c,d)  U(a^i)\nonumber,  \\
V(b_i) W(c,d)  &= W(c,d)  V(b_i).\label{19}
\end{align}
As a result, the relations in (\ref{19}) produce CCR with noncommutative positions using the quantization map, $\hat{\quad} ̂: P^{(A^i,B_i,C,D)}\rightarrow A^i \hat{p}^{\prime}_i+B_i\hat{q}^{i\prime}+C\hbar+D\theta$: 
\begin{align}\label{20}
[\hat{q}^i, \hat{q}^j]=i\theta^{ij}, \qquad [\hat{p}_i, \hat{p}_j]=0,\qquad [\hat{q}^i, \hat{p}_i]=i\hbar.
\end{align}
The Hilbert space of this representation are given by functions $\psi(q^i)\in L^2 (\mathbb{R}^2)$, then the corresponding unitary representation: 
\begin{align}\label{21}
U(a^i)\psi(q^i)&=   \psi(q^i-a^i),
\\
V(b_i)\psi(q^i)&= e^{iB_iq^i} \psi(q^i-\frac{1}{2}\theta^{ij}b_j),\label{22}
\end{align}
and these correspond to
\begin{align}\label{23}
\hat{q}^{i\prime}\psi(q^i)&=\left(q^i-\frac{i}{2}\theta^{ij}\frac{\partial}{\partial q^j}   \right)   \psi(q^i),
\\
\label{24}
\hat{p}^{\prime}_i\psi(q^i)&=-i\hbar\frac{\partial}{\partial q^j} \psi(q^i).
\end{align}
\\
 
\begin{center}
	\textbf{CONCLUSION}
\end{center} 
       In conclusion, we have studied quantum theory on two-dimensional plane of the phase space with modified symplectic structure using canonical group quantization. The canonical group obtained is the deformed Heisenberg group, $H^2_{\theta}$ (see Proposition). This is followed by contruction of its unitary irreducible representation which reproduces quantization of the noncommutative plane (Moyal plane)$^{5,7}$.  

\begin{center}
\textbf{ACKNOWLEDGMENTS}
\end{center}
The work is supported by Ministry of Higher Education (MOHE) of Malaysia and Universiti Pendidikan Sultan Idris (UPSI) scholarship, and also PUTRA grant, GP-IPM/2016/9473100.
\begin{center}
\textbf{REFERENCES}
\end{center}
\begin{spacing}{0.5}
\begin{enumerate}
  \item C. J. Isham, “Topological and global aspects of quantum theory”. in \textit{Relativity, Groups and Topology}, edited by B. S. Dewitt and R. Stora, North-Holland: Amsterdam, 1984, pp. 1059-1290.
  \item C. J. Isham and A. Kakas, \textit{Class.Quant.Grav.} \textbf{1}(16): 621 (1984).
  \item 	C. J. Isham and N. Linden, \textit{Class.Quant.Grav.} \textbf{5}(1): 71 (1988).
  \item H. Zainuddin, \textit{Phys.Rev.D.} \textbf{40}(2): 636 (1989).
  \item L. Gouba, Int. \textit{J.Mod.Phys.A.} \textbf{31}(19): 1630025 (2016)
  \item J. M. Romero, J. Santiago and J. D. Vergara, \textit{Phys.Lett.}A 310(1): 9-12 (2003)
  \item J. Gamboa, M. Loewe and J. Rojas, \textit{Phys.Rev.D} \textbf{64}(6): 067901 (2001).
  \item A. Ngendakumana, J. Nzotungicimpaye and L. Todjihounde, \textit{J. Math. Phys.} \textbf{55}(1): 013508 (2014)
  \item G. W. Mackey, \textit{Induced representations and quantum mechanics}, W. A. Benjamin, New
  York, (1968).
\end{enumerate}
\end{spacing}
\end{document}